# Silicon Oxide Electron-Emitting Nanodiodes

Gongtao Wu,[1] Zhiwei Li,[1] Zhiqiang Tang,[1] Dapeng Wei,[2] Gengmin Zhang,[1] Qing Chen,[1] Lian-Mao Peng,[1] Xianlong Wei [1,*]

[1] Key Laboratory for the Physics and Chemistry of Nanodevices and Department of Electronics, Peking University, Beijing 100871, China

[2] Chongqing Key Laboratory of Multi-scale Manufacturing Technology, Chongqing Institute of Green and Intelligent Technology, Chinese Academy of Sciences, Chongqing 400714, China

*Correspondence to: weixl@pku.edu.cn

**ABSTRACT:** Electrically driven on-chip electron sources that do not need to be heated have been long pursued because the current thermionic electron sources show the problems of high power consumption, slow temporal response, bulky size, etc., but their realization remains challenging. Here we show that a nanogap formed by two electrodes on a silicon oxide substrate functions as an electron-emitting nanodiode after the silicon oxide in the nanogap is electrically switched to a high-resistance conducting state. A nanodiode based on graphene electrodes can be turned on by a voltage of ~7 V in ~100 ns and show an emission current of up to several microamperes, corresponding to an emission density of ~$10^6$ A cm$^{-2}$ and emission efficiency as high as 16.6%. We attribute the electron emission to be generated from a metal-insulator-metal tunneling diode on the substrate surface formed by the rupture of conducting filaments in silicon oxide. An array of 100 nanodiodes exhibits a global emission density of 5 A cm$^{-2}$ and stable emission with negligible current degradation over tens of hours under modest vacuum. The combined advantages of a low operating voltage, fast temporal response, high emission density and efficiency, convenient fabrication and integration, and stable emission in modest vacuum make silicon oxide electron-emitting nanodiodes a promising on-chip alternative to thermionic emission sources.

**Main Text:** An electrically driven electron source is indispensable for many important electronic devices and instruments that rely on the motion of electrons in vacuum or rarefied gas (e.g., X-ray tubes, microwave tubes, electron microscopes and accelerators). To date, most of these applications still employ thermionic electron sources that were first developed more than a century ago.[1, 2] Because of the necessity to heat the emitter to the required high temperature (usually more than 1000 K), thermionic electron sources have associated problems such as high power consumption, distinct time delay in turning on/off, bulky size and unsatisfactory lifetime, substantially limiting their applications.[3] Alternatives to thermionic emission sources that do not require heating are therefore highly desirable, especially those with micro/nanoscale dimensions that can be fabricated and integrated on a chip using semiconductor microfabrication technologies for scaling down vacuum electronic devices.[4, 5] Previous efforts in this area include tip-based field emission sources,[6, 7] metal (M)-insulator (I)-metal (M) multilayer tunneling emission sources,[8, 9] reverse-biased P-N junction sources,[10, 11] forward-biased Schottky barrier sources,[12, 13] and negative-electron-affinity sources.[14, 15] Field emission sources require ultra-high vacuum to achieve stable emission with long lifetime and high operating voltage (usually larger than 100 V).[4] Other alternatives encountered problems of either operational instability and short lifetime when the surface was cesiated[16] or low emission density and efficiency in the absence of a cesiated surface.[9, 11, 13, 15] Despite more than a half century of research into these alternatives, an electrically driven on-chip electron source with low operating voltage and dense, efficient and stable emission in a modest vacuum for practical applications is still unavailable. In this paper, we report a silicon oxide electron-emitting nanodiode (EEND) with a simple structure and exciting electron emission (EE) characteristics that are promising for vacuum electronic device applications.

A silicon oxide EEND is simply a nanogap spaced by two electrodes on a silicon oxide substrate. A typical structure of a silicon oxide EEND based on graphene electrodes is shown in **Figure 1**a-b, where two graphene films form an 88-nm-wide gap on the surface of a thermal-oxidized Si wafer with a 300-nm silicon oxide surface layer (Supplementary Figure S1). We found that considerable EE from the nanogap could occur after electroforming of silicon oxide in the nanogap (Supplementary Figure S2). Silicon oxide in the nanogap was electroformed by ramping up bias voltage applied to the electrodes until a sudden steep increase of conduction current, indicating that pristine insulating silicon oxide was transformed to a conducting state. The emission current from the EEND was measured by an electrode located approximately 50 μm above the diode and applied with a voltage of 200 V. Figure 1c shows the simultaneously measured transport and EE characteristics of the EEND based on graphene electrodes after the silicon oxide in the nanogap was electroformed. Electroformed silicon oxide in the nanogap exhibited unipolar resistive switching (RS) behavior. A sudden switching from a high-resistance ($R$) state with $R$=45.6 MΩ to a low-R one with $R$=42.6 kΩ was observed at a bias voltage ($V$) of 2.6 V, and reversal switching from a low-R state with $R$=9.5 kΩ to a high-R one with $R$=96.9 kΩ was achieved at $V$=6.9 V. While the RS behavior of electroformed silicon oxide in the nanogaps has been intensively reported,[17-22] here we report the observation of EE associated with RS of silicon oxide. At the moment of RS from a low-R state to a high-R one at $V$=6.9 V, the emission current suddenly appeared and then increased exponentially with $V$ reaching 3.2 μA at $V$=21.3 V. Considering the conduction current of 103.8 μA at $V$=21.3 V, the emission efficiency (the ratio of emission current to conduction current) of 3.1% was obtained. The emission efficiency increases with the bias voltage and could reach up to 16.6% at $V$=40 V (Supplementary Figure S3), higher than the record efficiency (11%) of an M-I-M multilayer tunneling emission source.[9] The sudden emergence of the emission current at the moment of RS at $V$=6.9 V indicates that the observed EE was likely caused by the RS of silicon oxide from a low-R state to a high-R state.

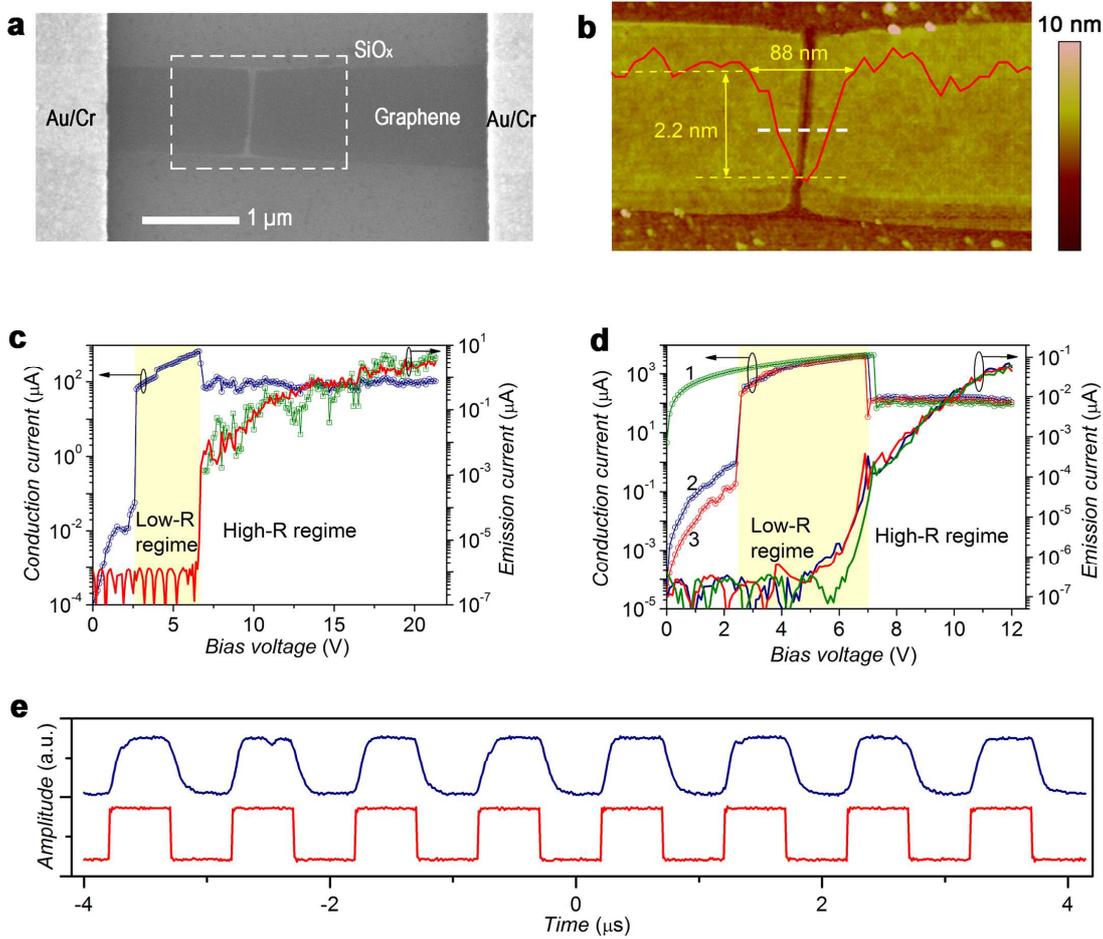

**Figure 1. Structure and performances of single EENDs based on graphene electrodes.** (a) Scanning electron microscopy (SEM) image of a silicon oxide EEND based on graphene electrodes. (b) Atomic force microscopy (AFM) image of the framed area in (a), indicating the formation of an 88-nm-wide nanogap spaced by graphene electrodes. Red solid line indicates the height profile along the white dashed line. (c) Simultaneously measured transport characteristics (blue curve) and EE characteristics (red curve) of the EEND. The calculated emission characteristics based on Equation 1 are shown in olive symbols. (d) Transport characteristics and EE characteristics of the EEND in three different initial resistance states, as indicated by curves 1, 2 and 3. The transport characteristics and emission characteristics displayed in the same color were measured simultaneously. RS results in low-R and high-R regimes in (c) and (d) with the low-R regime highlighted by the yellow background. (e) Temporal response (blue curve) of electron emission from the EEND when it was driven by a square wave voltage (red curve).

The resistance of electroformed silicon oxide in a nanogap showed memory of the voltage-application history and could be set to different resistance states.[17, 19, 22] Figure 1d shows the transport and EE characteristics of the EEND when it was set to three different resistance states, as shown by curves 1, 2 and 3. The three transport curves show different resistance states at low bias voltage and converge to the same resistance states at high bias voltage due to the RS at 2.4 V. Nearly overlapped EE characteristics were obtained for the three different resistance states, especially in the high-R regime after RS at $V=\sim7.0$ V. The EE characteristics of the EEND therefore exhibited no dependence on the initial resistance states of silicon oxide in the nanogap at low bias voltage. The overlapped EE characteristics in the high-R regime agree with our deduction that EE from the EEND in the high-R regime is caused by the RS from a low-R state to a high-R one, as the RS with nearly the same initial low-R and final high-R states was observed at $V=\sim7.0$ V for the three curves.

Unlike Figure 1c where electron emission was only observed in the high-R regime after the RS, a small emission current of 0.2-0.3 nA was also observed in the low-R regime before the RS at $V=\sim7.0$ V for the EEND shown in Figure 1d. Abrupt changes of the emission current and its increasing slope with $V$ were observed at the moment of the RS, indicating different mechanisms responsible for the EE before and after the RS. Electron emission before RS is attributed to thermionic emission from graphene electrodes under Joule heating,[23] as a thermionic emission current of a similar magnitude was observed from the same graphene film under Joule heating of a conduction current similar to that before RS at $V=\sim7.0$ V in Figure 1d (Supplementary Figure S4). Thermionic emission from the graphene electrodes was only observed for the EENDs with a large conduction current density of more than $\sim1.5$ mA $\mu m^{-1}$.

To elucidate the origin of the observed considerable EE in the high-R regime, we carried out additional experiments. First, we fabricated nanogaps spaced by graphene electrodes on a silicon nitride substrate, which was demonstrated to show no RS behavior, [17, 18] following the same procedures as for the nanogaps on the silicon oxide substrate and found that these were insulating without electroforming, and no emission current was measured up to a high $V$ of 100 V (Supplementary Figure S5). This enables us to exclude field emission between the graphene electrodes as a possible mechanism responsible for our observed considerable EE in the high-R regime. Second, we fabricated ~90-nm-wide nanogaps spaced by metal electrodes (70 nm Au/5 nm Cr) on both silicon oxide and silicon nitride substrates (Supplementary Figure S6). The nanogaps on the silicon oxide substrate were initially insulating, and no emission current was measured until the silicon oxide in the nanogaps was electroformed at a $V$ of 35 V (Supplementary Figure S6b). After electroforming, the silicon oxide in the nanogap became conductive with the RS behavior, and the emission current became measurable even if $V$ was much smaller than 35 V used for electroforming, namely, larger than ~8 V (Supplementary Figure S6c). In contrast, the nanogaps spaced by metal electrodes on the silicon nitride substrate remained insulating, and no emission current was detected up to a high $V$ of 210 V (Supplementary Figure S6e). The comparative results for the nanogaps based on different electrodes (graphene *vs* metal) and different substrates (silicon oxide *vs* silicon nitride) unambiguously indicate that the EE in the high-R regime of Figure 1c-d originated from the electroformed silicon oxide in the nanogap.

The temporal response of the EE arising from the silicon oxide in the nanogap spaced by graphene electrodes was tested by driving the EE using a square wave voltage and detecting the emitted electrons using an Everhart-Thornley detector (ETD).[23] Figure 1e displays the output of the ETD detector together with the input square wave voltage. The low and high voltage levels of the square wave voltage were set to 11 and 14.7 V, respectively, in order to be smaller and larger than the RS voltage of the silicon oxide in the nanogap from the high-R state to the low-R one. The turn-on and -off times (defined as the time to increase to 90% or decrease to 10% of the peak current) of the EE were determined to be approximately 110 and 130 ns, respectively. These values are in good agreement with the time (~50 ns) of the switch of the silicon oxide in the nanogap from the low-R state to the high-R state and the time (~100 ns) of the reversal switch.[21] The good agreement provides additional evidence that the EE from the silicon oxide in the nanogaps is caused by the RS of silicon oxide from the low-R state to the high-R state.

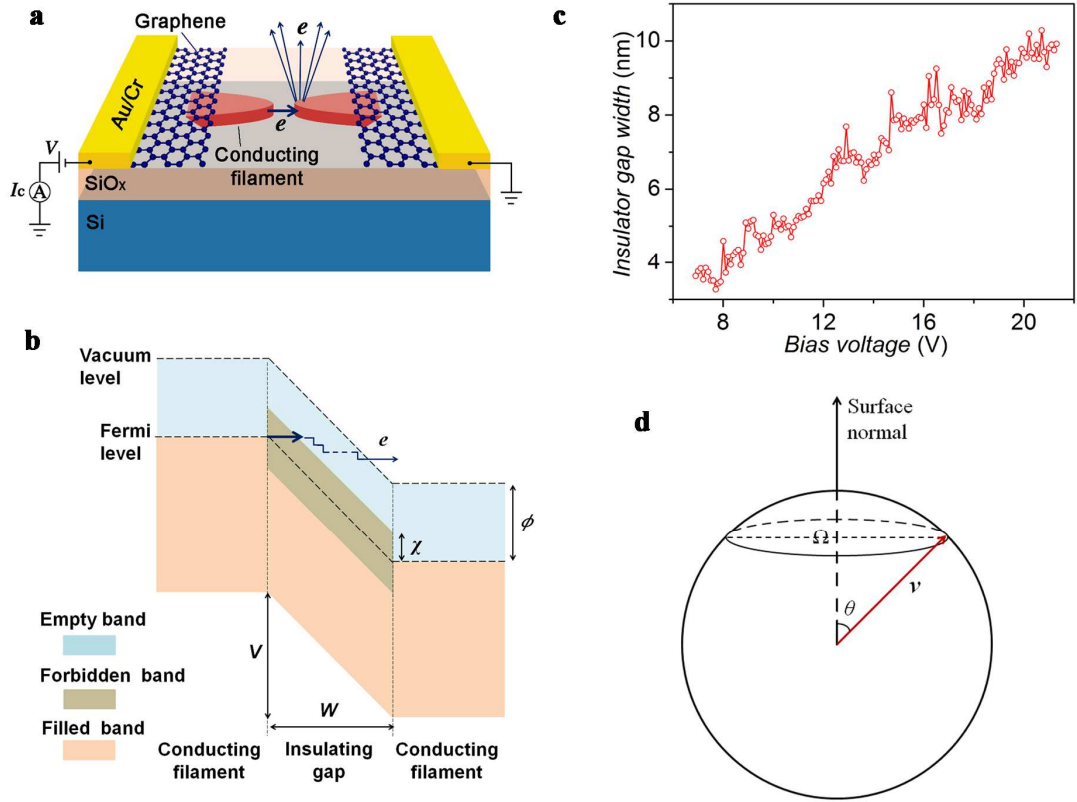

**Figure 2. Mechanism of electron emission from the silicon oxide EENDs.** (a) Schematic diagram showing electron emission from a surface M-I-M tunneling diode formed by the rupture of the conducting filament in a silicon oxide nanogap spaced by graphene electrodes. (b) Schematic energy band diagram showing electron transport in an M-I-M tunneling diode. (c) Calculated width of insulator gap at different bias voltage for the EEND in Figure 1c. (d) A diagram showing emission probability ($\Omega/4\pi$) of electrons with kinetic energy of $\frac{1}{2}mv^2 > E_f + \phi$. Note: $V$, bias voltage; $I_c$, conduction current; $W$, width of insulating gap; $\chi$, tunneling barrier at M-I interface; $\phi$, work function; →, electron transport paths.

We attribute the EE in the high-R regime of Figure 1c-d to the M-I-M tunneling diode on the substrate surface formed by the rupture of the conducting filament in silicon oxide. Conducting filaments made of semi-metallic silicon were found to form in silicon oxide during electroforming through a $SiO_x \rightarrow Si$ electrochemical reduction process.[18, 20] The reversible point rupture and reconnection of conducting filaments embedded in silicon oxide was proposed to be responsible for the reversible RS between the low-R and high-R states. [18-20] As a result, a metal (semi-metallic silicon)-insulator (silicon oxide)-metal (semi-metallic silicon) tunneling diode, as schematically shown in **Figure 2**a, forms at the moment of the RS from the low-R state to the high-R state due to the rupture of the conducting filament. When a bias voltage is applied to the tunneling diode, electrons will tunnel from the metallic region (conducting filament) to the insulating region (silicon oxide) and are accelerated in the latter. If the bias voltage is larger than the work function of the metallic region, electrons can be accelerated to above the vacuum level when they enter the opposite metallic region. Conducting filaments in electroformed silicon oxide were found to form within a shallow region with a ~ 4-nm depth from the substrate surface;[22] this is smaller than the inelastic mean free path of 6 nm in silicon, [24] and therefore, the electrons that are accelerated to above the vacuum level will be possibly emitted when they move toward the substrate surface after being elastically scattered in the opposite metallic region. Compared with vertical multilayer M-I-M tunneling emission sources that are fabricated by depositing metallic and insulating materials layer by layer, [8, 9] the surface M-I-M tunneling diode is electrically formed in silicon oxide by taking advantage of the RS of silicon oxide, which endows the EENDs with the properties of easy fabrication and electrical modulation of the diode structures.

Conducting filaments in electroformed silicon oxide in the nanogap were found to be highly localized to a size of less than 10 nm, [18, 20, 21] indicating that electrons were probably emitted from a highly localized region in the nanogaps. To confirm this, nanogaps spaced by carbon nanotube (CNT) electrodes were fabricated on a silicon oxide substrate. Silicon oxide in the nanogap spaced by 18-nm-wide CNT electrodes exhibited similar RS and EE characteristics to those presented in Figure 1c after electroforming, and an emission current of up to 2.8 μA was measured (Supplementary Figure S7). After electrons enter the opposite metallic region, they should be emitted before any impact ionization that can cause considerable energy loss. Electrons are therefore emitted at a distance comparable to the mean free path for impact ionization after entering the opposite metallic region. Considering the conducting filament width of 18 nm and the mean free path for impact ionization of 19 nm, [24] the emission current of 3.2 μA in Figure 1c corresponds to an emission density of $0.94\times10^6$ A cm$^{-2}$, comparable to that of field emission from a sharp tip. [7]

To quantitatively describe the EE from the EEND, a model based on the above picture of the M-I-M tunneling diode was proposed to calculate the emission current of the EEND from its conduction current. An energy band diagram illustrating electron tunneling in an M-I-M diode is schematically shown in Figure 2b. The increase of resistance with bias voltage after RS in Figure 1c-d was attributed to the widening of insulator gap in an M-I-M diode due to voltage stress. The width of insulator gap at different bias voltage was firstly calculated by fitting the conduction current after RS with Fowler-Nordheim law. Assuming a triangular barrier and a temperature of 0 K, the conduction current of an EEND after RS is described by Fowler-Nordheim law: $I_c = A\dfrac{q^3 E^2}{8\pi h \chi}\exp(-\dfrac{4(2m)^{1/2}\chi^{3/2}}{3\hbar q E})$, [25] where $h$ is Planck constant, $q$ is electron charge, $m$ is electron mass, $A$ is the end face area of ruptured conducting filament, $\chi$ is tunneling barrier at the M-I interface, $E=V/W$ is strength of electric field with $V$ being bias voltage and $W$ being the width of insulator gap. The end faces of the ruptured conducting filament are assumed to be rectangular with dimensions of 4 nm ×10 nm independent of bias voltage, according to the reported depth (4 nm) and width (~10 nm) of conducting filaments in silicon oxide.[18,22] Considering a tunneling barrier ($\chi$) of 1.0 eV, the width $W$ of insulator gap for the EEND in Figure 1c was found to increase linearly with bias voltage from 3.8 to 10.7 nm when bias voltage increased from 6.9 to 21.3 V (Figure 2c).

After tunneling into insulator gap, electrons are accelerated by electric field and at the same time encounter scattering of phonons. When electrons reach the M-I interface, their energy distribution can be approximately described by the Gaussian distribution:

$$D(\varepsilon) = \frac{1/\hbar\omega}{(2\pi L/\lambda)^{1/2}} \exp(-\frac{[(\hbar\omega L/\lambda)-(E_f+V-\varepsilon)]^2}{(\hbar\omega)^2(2L/\lambda)})$$ ,[26] where $\varepsilon$ is electron energy with respect to the bottom of conduction band in metallic region, $E_f$ is Fermi energy, $\lambda$ is mean free path of electrons, $\hbar\omega$ is phonon energy, $L=W-\chi/E$ is electron transport distance in insulator gap after tunneling. To calculate emission probability of electrons after they enter the opposite metallic region, we assume that the moving of the electrons is isotropic as a result of elastic scattering before they encounter considerable energy loss. Only those electrons that moving toward the substrate surface and having a kinetic energy normal to the surface higher than the surface barrier can be emitted, so the emitted electrons have to meet the criterion $\frac{1}{2}m(v\cos\theta)^2 > E_f+\phi$, where $v$ is electron velocity, $\theta$ is the angle between electron moving direction and surface normal (Figure 2d), $\phi$ is the work function of metallic region. In other words, for an electron with kinetic energy $\frac{1}{2}mv^2 > E_f+\phi$, it can be emitted from the substrate surface only when it distributes in the solid angle of $\Omega = 2\pi[1-(\frac{E_f+\phi}{mv^2/2})^{1/2}]$. Here $\Omega$ is the solid angle corresponding to the angle of $\theta = \cos^{-1}(\frac{E_f+\phi}{mv^2/2})^{1/2}$ (Figure 2d). Since $\varepsilon$ was defined with respect to the bottom of conduction band, we have $\varepsilon = \frac{1}{2}mv^2$. In the case of electron moving isotropically before emission, we get the emission probability $P(\varepsilon) = \Omega/4\pi = [1-(\frac{E_f+\phi}{\varepsilon})^{1/2}]/2$ for electrons with energy $\varepsilon > E_f+\phi$. After obtaining the electron energy distribution and emission probability, the relation between emission current ($I_e$) and conduction current ($I_c$) of an EEND can be written as:

$$I_e = I_c \frac{\int_{E_f+\phi}^{\infty} D(\varepsilon)P(\varepsilon)d\varepsilon}{\int_{-\infty}^{\infty} D(\varepsilon)d\varepsilon} \tag{1}$$

The EE characteristics of the EEND in Figure 1c were well reproduced by the model with reasonable parameter values of $\lambda$=0.134 nm, [26] $\hbar\omega$ =0.22 eV, $\phi$=5.0 eV, $E_f$=0.77 eV, despite the frequent fluctuations in the calculated emission current (as shown by olive symbols in Figure 1c). The frequent fluctuations can be explained by the fluctuations in the conduction current and the irregular evolution of the real end faces of ruptured conducting filaments with the bias voltage.

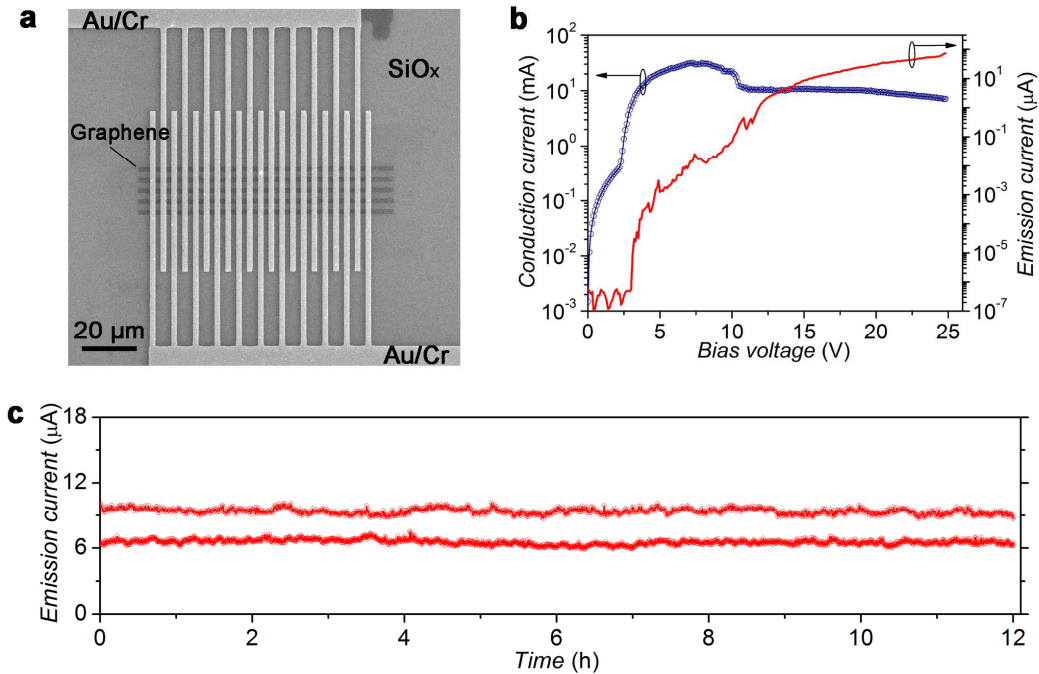

**Figure 3. Structure and performances of the EEND array.** (a) SEM image of a 5×20 parallel array of EENDs based on graphene electrodes. (b) Simultaneously measured transport characteristics and emission characteristics of a 5×20 parallel EEND array. (c) Long-term stability of the emission current from a 100 EEND array under fixed bias voltages of 18 and 21 V and a vacuum of ~5×10$^{-6}$ Pa.

The simple structure of the EENDs and the wide use of the silicon oxide substrate in microfabricated electronic devices enable the convenient integration of EENDs using microfabrication technologies. **Figure 3**a shows a 5×20 array of EENDs based on graphene electrodes integrated in parallel through interdigital electrodes. The EEND array exhibits RS and EE performances similar to those of the single EEND in Figure 1c-d (Figure 3b). An emission current of 73.4 µA was obtained at $V$=24.8 V. Considering the effective area of 82

μm×18 μm in the EEND array, a global emission density of 5 A cm$^{-2}$ was obtained, comparable to that of a practical high-current-density thermionic emission source. [2] Figure 3c shows the long-term stability of the emission current from an array of 100 EENDs, which exhibited an emission current degradation of <3% and a root-mean-square noise ratio $\langle \Delta I^2 \rangle^{1/2}/\langle I \rangle$ of <4% in 24 hours under a modest vacuum of ~5×10$^{-6}$ Pa.

In summary, by electrically forming a surface M-I-M tunneling diode in silicon oxide, a nanogap spaced by two electrodes on a silicon oxide substrate functions as an electron-emitting nanodiode with low operating voltage, fast temporal response, and dense, efficient and stable emission in a modest vacuum. The outstanding electron emission performances and the convenient fabrication and integration by using microfabrication technologies make the EEND a promising on-chip electron source in the forms of both single nanoscale sources for micro/nanoscale vacuum electronic devices and large scale source arrays to replace thermionic electron sources in conventional vacuum electronic devices. The on-chip nature of EENDs also provides a new route of scaling down vacuum electronic devices on a chip.

**Experimental Section**
*Device fabrication.* Electron-emitting nanodiodes (EENDs) were fabricated using microfabrication technologies. Graphene films were grown by chemical vapor deposition (CVD) method, [27] and were transferred to a silicon oxide or silicon nitride substrate by using PMMA (polymethyl methacrylate) (see Supplementary Figure S8 for the characterizations of graphene). [28] Silicon oxide substrates were purchased from Silicon Quest International (SQI) with 300 nm silicon oxide surface layer on n+ doped silicon wafers. Silicon nitride substrates were prepared by growing silicon nitride film on silicon wafers at a temperature of 130 °C and using SiH$_4$ and NH$_3$ as reactive gases by inductively coupled plasma CVD method. XPS characterizations of silicon oxide and silicon nitride substrates are shown in Supplementary Figure S1. The thickness of silicon nitride film was measured to be 492 nm by ellipsometry.

After a graphene film was transferred to a silicon oxide or silicon nitride substrate, it was tailored into ribbons in pre-designed dimensions and configurations by electron beam lithography (EBL) followed by plasma etching. Metal electrodes (70 nm Au/5 nm Cr) contacting on graphene ribbons were then fabricated using standard EBL, metal film deposition, and lift-off processes. After the device structure with a graphene film connected between two metal electrodes was fabricated, the graphene film was broken down by ramping up electrical voltage applied to it until a sudden steep drop of the conduction current (Supplementary Figure S4). The breakage of a graphene film into two sections under the extreme electrical and/or thermal stress results in a nanogap throughout its width, and a nanogap spaced by two graphene electrodes was obtained.

*Electron emission measurement.* To measure emission current from an EEND or EEND array, an electrode applied with a 200 V voltage was placed ~50 μm above the diode to collect the emitted electrons (Supplementary Figure S9a). To facilitate electron collection by the top electrode, a negative voltage was applied to one electrode of EENDs with the other electrode grounded to drive electron emission, and a voltage of 50 V was applied to the bottom Si layer to prevent electron collection by side electrodes. Emission current from single EENDs was measured in a Janis probe station with a vacuum level of $\sim 10^{-3}$ Pa and a tungsten probe with a diameter of 30 μm was used as the top collecting electrode. Emission current from EEND arrays was measured in a home-made vacuum chamber with a vacuum level of $\sim 5\times 10^{-6}$ Pa and a metal plate with a diameter of 10 mm was used as the top collecting electrode. The voltage and current was sourced and measured by using a Keithley 4200 semiconductor characterization system. The temporal response of EENDs was measured in the chamber of a SEM (FEI Quanta 600F) by driving the electron emission using a square wave voltage and detecting electron emission using the Everhart-Thornley detector (ETD) of the SEM (Supplementary Figure S9b). The vacuum in SEM chamber was $\sim 5\times 10^{-3}$ Pa. Square wave voltage was applied through a signal generator (Agilent 33220A). The input square wave voltage and the output signal of ETD were recorded by an oscilloscope (Aglient DSO7054A). Electrical connection to EENDs was achieved using nanoprobes (Kleindiek MM3A) installed in the SEM chamber. When measuring the temporal response of EENDs, electron beam of SEM was blanked and the gain of ETD was adjusted to avoid overflow of the detector during measurements. The grid voltage and scintilator voltage of ETD were set to be 210 and 10000 V, respectively.

**Acknowledgements**
We thank Y. T. Song and S. M. Chen at the Institute of Process Engineering, Chinese Academy of Science for XPS characterization, and S. Gao at Peking University for AFM characterization. This work was supported by the National Key Research and Development Program of China grant 2017YFA0205003, the National Basic Research Program of China grant 2013CB933604, and the National Nature Science Foundation of China grants 61371001 and 61621061.

**Author contributions**
X.W. and G.W. conceived the project, designed the experiments and analyzed the data; X.W. supervised the project, proposed the emission mechanism and model, and wrote the manuscript; G.W., Z.L., and Z.T. performed device fabrication and characterization; D.W. grew graphene; G.Z., Q.C. and L.-M.P. built research infrastructures and contributed considerable suggestions.

# Supporting Information

**Silicon Oxide Electron-Emitting Nanodiodes**

*Gongtao Wu, Zhiwei Li, Zhiqiang Tang, Dapeng Wei, Gengmin Zhang, Qing Chen, Lian-Mao Peng, Xianlong Wei\**

1. Characterization of silicon oxide and silicon nitride substrates

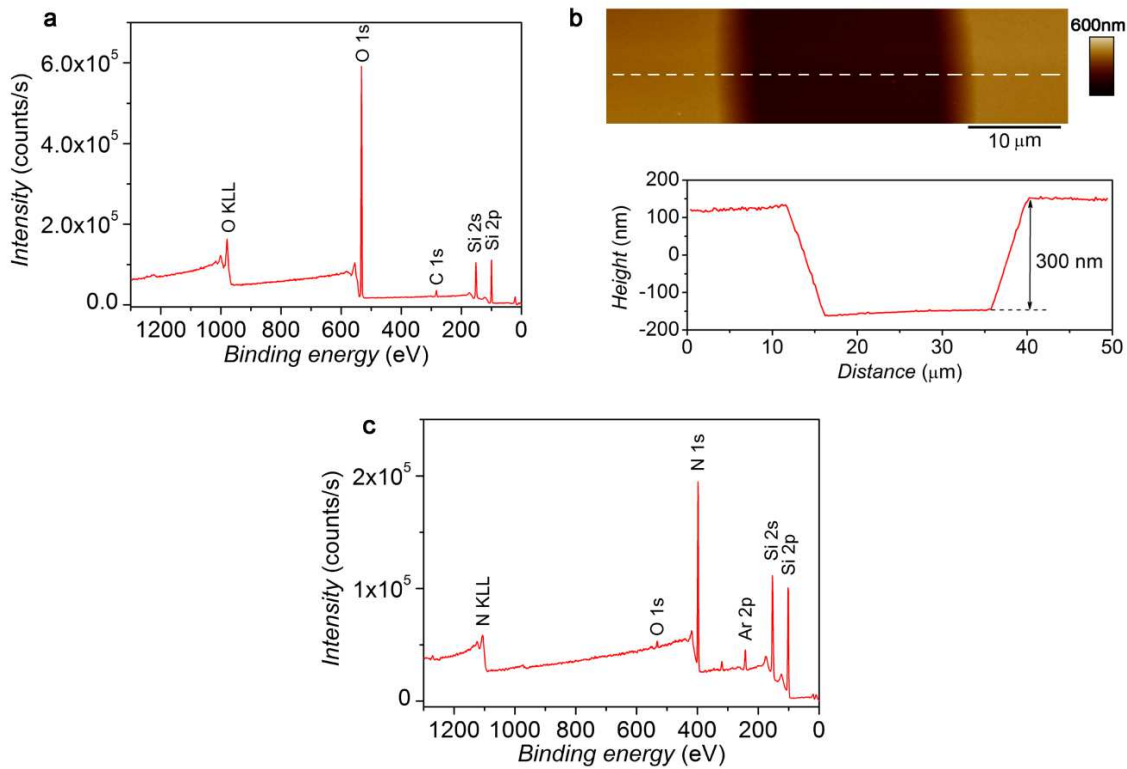

Figure S1. Characterization of silicon oxide and silicon nitride substrates. (a) XPS spectrum of silicon oxide substrate showing characteristic Si 2p peak (103.3 eV) and O 1s peak (532.7 eV). (b) Upper: AFM image of a trench on silicon oxide substrate with silicon oxide in the trench etched; bottom: height profile along the dashed line in the upper image. (c) XPS spectrum of silicon nitride substrate showing characteristic Si 2p peak (101.7 eV) and N 1s peak (397.5 eV). The XPS spectrum of silicon nitride substrate was obtained after ~15 nm surface layer was removed by Ar ion beam.

Chemical compositions of silicon oxide and silicon nitride substrates used in this work were characterized by using x-ray photoelectron spectroscopy (XPS). A XPS spectrum of the silicon oxide substrate is shown in Figure S1a, where characteristic Si 2p peak at 103.3 eV and O 1s peak at 532.7 eV for silicon oxide can be identified. Quantification analysis of the XPS spectrum gives a composition of $SiO_x$ with $x=1.82$. The 300 nm thickness of silicon oxide layer was confirmed by AFM height profile of a trench on the substrate with silicon oxide in the trench was removed by chemical etching (Figure S1b). A XPS spectrum of silicon nitride substrate is shown in Figure S1c, where characteristic Si 2p peak at 101.7 eV and N 1s peak at 397.5 eV for silicon nitride can be identified. Quantification analysis of the XPS spectrum gives a composition of $SiN_x$ with $x$=0.9. The thickness of silicon nitride film was measured to be 492 nm by ellipsometry.

## 2. Electroforming of silicon oxide in nanogaps

After a nanogap spaced by two electrodes was fabricated on silicon oxide substrate, it was electroformed by ramping up bias voltage applied to the electrodes until a sudden steep increase of conduction current. Figure S2a shows the simultaneously measured conduction current and emission current of a nanogap spaced by graphene electrodes until it was electroformed at 14 V. Before electroforming, silicon oxide in the nanogap was insulating and no emission current was measured up to a bias voltage of 14 V. After electroforming, silicon oxide in the nanogap exhibited unipolar RS behaviors and significant emission current was measured at a comparatively low voltage of ~7 V (Figure 1c). Some devices were observed to be spontaneously electroformed at the moment of graphene film breakage [1], and no further electroforming process was needed, as shown by an example in Figure S2b. When graphene film was broken down at 6.3 V, conduction current dropped down steeply from ~7 mA to 65 µA, but not completely disappeared. The residual electrical conduction after graphene breakage is attributed to the conduction of electroformed silicon oxide in the nanogap formed by graphene breakage. This is supported by larger emission current measured after graphene breakage as compared to thermionic emission current from graphene film before graphene breakage. Emission current becomes highly fluctuating after graphene breakage and has a smaller increasing rate with bias voltage as compared to thermionic emission current before graphene breakage, in good agreement with emission current from electroformed silicon oxide as shown in Figure 1d. The electrical conduction after electroforming was attributed to the forming of conducting filaments in silicon oxide [2, 3].

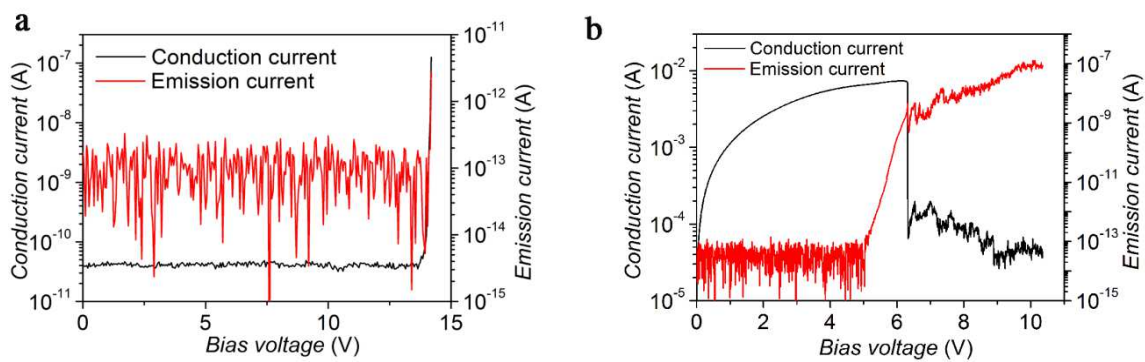

**Figure S2. Electroforming of silicon oxide in a nanogap spaced by graphene electrodes.** (a) Simultaneously measured transport characteristics and emission characteristics of a silicon oxide nanogap until it was electroformed at 14 V, where a sudden steep increase of conduction current and emission current was observed. To make silicon oxide in a nanogap electroformed, bias voltage applied to the electrodes was gradually increased until a sudden steep increase of conduction current. (b) Simultaneously measured transport characteristics and emission characteristics of a silicon oxide nanogap, showing the spontaneous electroforming of silicon oxide at the moment of graphene breakage at 6.3 V.

## 3. An EEND with emission efficiency up to 16.6% at 40 V

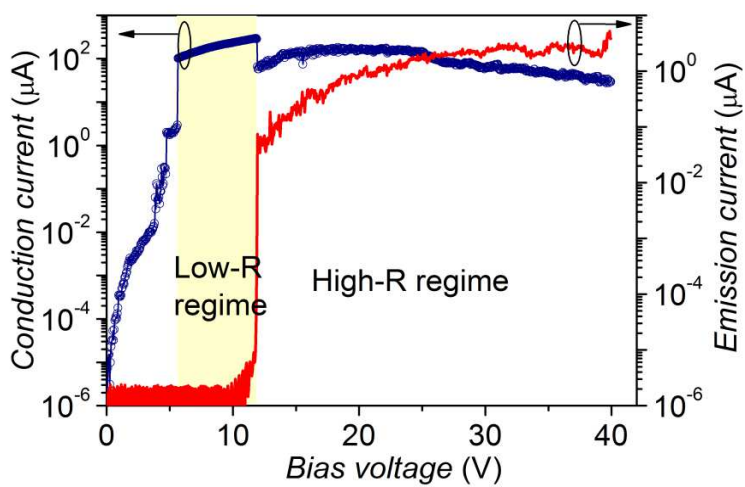

**Figure S3. Performances of an EEND measured up to a bias voltage of 40 V.** Simultaneously measured transport and emission characteristics of an EEND based on graphene electrodes. The EEND exhibited RS from a low-R state to a high-R one at 11.9 V. At a bias voltage of 40 V, an emission current of 5.2 µA and a conduction current of 31.3 µA were obtained, which corresponds to an emission efficiency (the ratio of emission current to conduction current) of as high as 16.6%.

## 4. Thermionic electron emission from graphene electrodes

When fabricating nanogaps by electrically breaking down graphene films (Figure S4a,b), electron emission was observed prior to the breakage of graphene film. Figure S4c shows the simultaneously measured conduction current and emission current until the breakage of the graphene film when fabricating the EEND in Figure 1d. Emission current became measurable at $V$=6.9 V and then increased exponentially with bias voltage to 0.3 nA until it dropped suddenly to zero (below the measurement limit of 0.1 pA) at the moment of graphene breakage at $V$=8.1 V. The observed electron emission is attributed to thermionic electron emission from hot graphene film under the Joule heating according to our previous paper [4]. The magnitudes of emission current (0.3 nA) and conduction current (4.8 mA) prior to graphene breakage coincide well with those of the emission current (0.2-0.3 nA) and conduction current (4.7 mA) measured prior to RS at V=~7 V in Figure 1d. This provides direct evidence that emission current prior to RS at $V$=~7 V in Figure 1d was thermionic emission current from graphene electrodes.

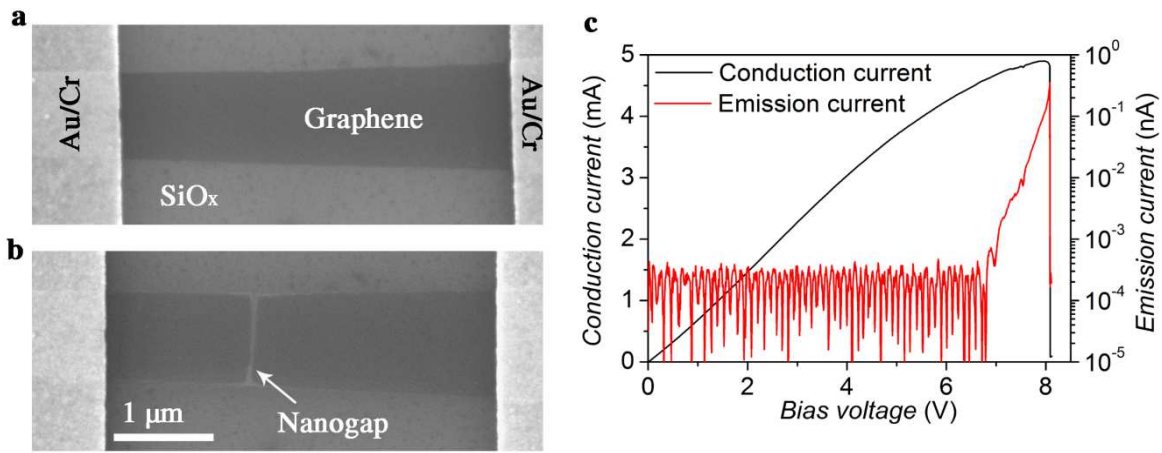

**Figure S4. Fabricating a nanogap spaced by graphene electrodes via electrical breakdown.** (a-b) SEM images of a graphene film before and after a nanogap was formed in the middle by breaking it through Joule heating. (c) Simultaneously measured conduction current and emission current until the breakage of graphene film at 8.1 V when fabricating the EEND in Figure 1d.

## 5. Nanogaps spaced by graphene electrodes on silicon nitride substrates

Nanogaps spaced by graphene electrodes were fabricated on silicon nitride substrate following the same processes as those on silicon oxide substrate. Figure S5a shows a SEM image of such a device, where a meandering slit formed across graphene film. In contrast with their counterparts on silicon oxide substrate, nanogaps spaced by graphene electrodes on silicon nitride substrate remained insulating up to a high bias voltage of 100 V without electroforming and RS, in agreement with previous reports [5,6], and no emission current was detected (Figure S5b).

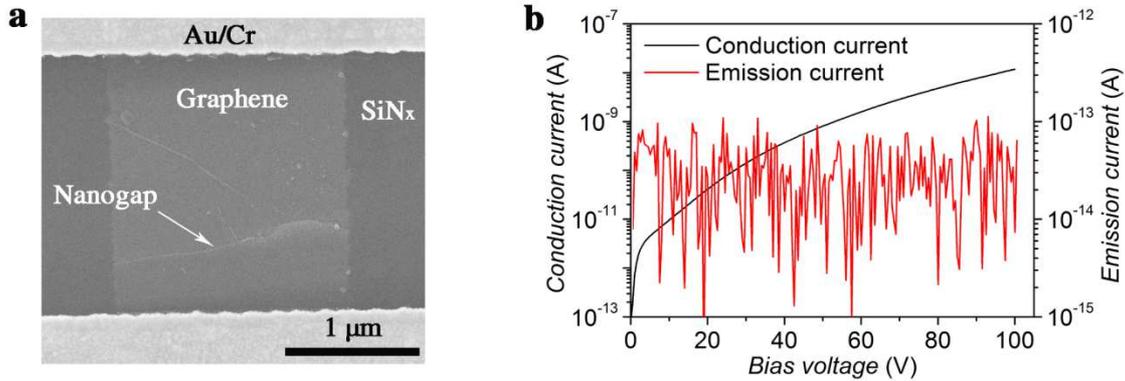

**Figure S5. Results of a nanogap spaced by graphene electrodes on silicon nitride substrate.** (a) SEM image of a nanogap spaced by graphene electrodes on silicon nitride substrate. (b) Simultaneously measured transport characteristics and emission characteristics of a nanogap spaced by graphene electrodes on silicon nitride substrate. The nanogap maintained insulating up to a high bias voltage of 100 V without electroforming and no emission current was detected.

## 6. Nanogaps spaced by metal electrodes

Nanogaps spaced by metal electrodes (70 nm Au/5 nm Cr) were fabricated on both silicon oxide and silicon nitride substrates. For a ~90 nm wide nanogap on silicon oxide substrate (Figure S6a), it was electroformed at a 35 V bias voltage as shown in Figure S6b, where a sudden increase of conduction current and emission current was observed at 35 V. Before electroforming at 35 V, silicon oxide was insulating as expected and no emission current was detected. The transport and emission characteristics of the nanogap measured after electroforming are shown in Figure S6c. It can be seen that, silicon oxide in the nanogap became electrical conducting and exhibited RS behaviors after electroforming. A transition from high-resistance (R) state to low-R one was observed at $V=\sim 3$ V with sudden current increase by a magnitude of two orders, and highly fluctuating conduction current showing a negative differential conductance was observed after $V=\sim 3$ V. The RS behaviors agree well with those reported previously [1,2]. Emission current became measurable after $V=\sim 7.5$ V after electroforming and increased with bias voltage. The emission current looks more fluctuating than that of nanogaps spaced by graphene electrodes in Figure 1c-d, probably caused by its more fluctuating conduction current. Moreover, emission current from the nanogap spaced by metal electrodes is much smaller than that of nanogaps spaced by graphene electrodes in Figure 1c-d. This can be explained by the fact that a larger fraction of electrons emitted from silicon oxide in the gap are collected by the side thick metal electrodes (75 nm thick) as compared to the case of ultrathin graphene electrodes (< 1 nm thick). It is why we mainly focused on EENDs based on graphene electrodes in this paper. In contrast with the nanogap on silicon oxide substrate, a 90 nm wide nanogap on silicon nitride substrate remained insulating up to a high bias voltage of 210 V without electroforming and no emission current was detected (Figure S6d-e), agreeing with the results in Figure S5. The comparative experiments on silicon oxide and silicon nitride substrates unambiguously indicate that the observed electron emission originated from electroformed silicon oxide in nanogaps.

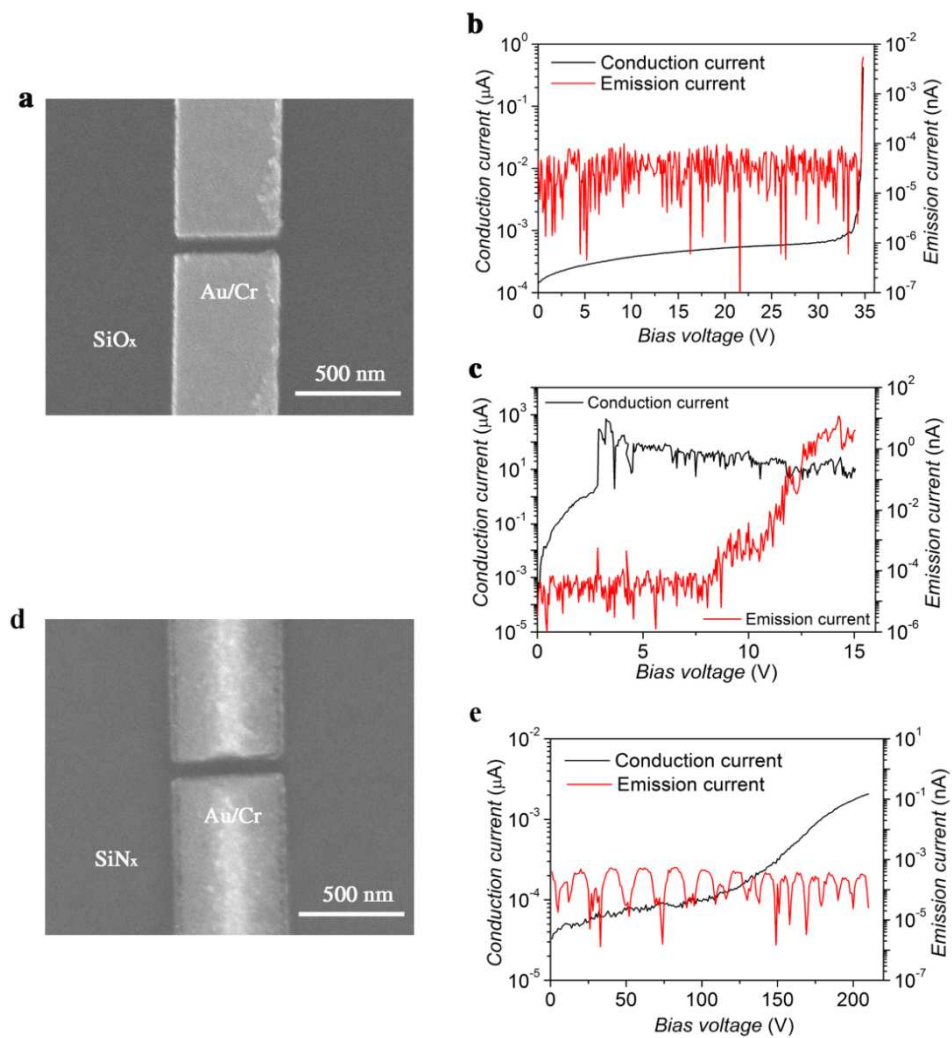

**Figure S6. Results of nanogaps spaced by metal electrodes (Au/Cr) on silicon oxide and silicon nitride substrates.** (a) SEM image of a 90 nm wide nanogap spaced by metal electrodes on silicon oxide substrate. (b-c) Simultaneously measured transport characteristics and emission characteristics of a silicon oxide nanogap spaced by metal electrodes before (b) and after (c) it was electroformed at 35 V. (d) SEM image of a 90 nm wide nanogap spaced by metal electrodes on silicon nitride substrate. (e) Simultaneously measured transport characteristics and emission characteristics of a silicon nitride nanogap spaced by metal electrodes. While electron emission from the silicon oxide nanogap was observed at ~8 V after electroforming, the silicon nitride nanogap remained insulating up to high bias voltage of 210 V without electroforming and no emission current was detected.

## 7. Silicon oxide EEND based on CNT electrodes

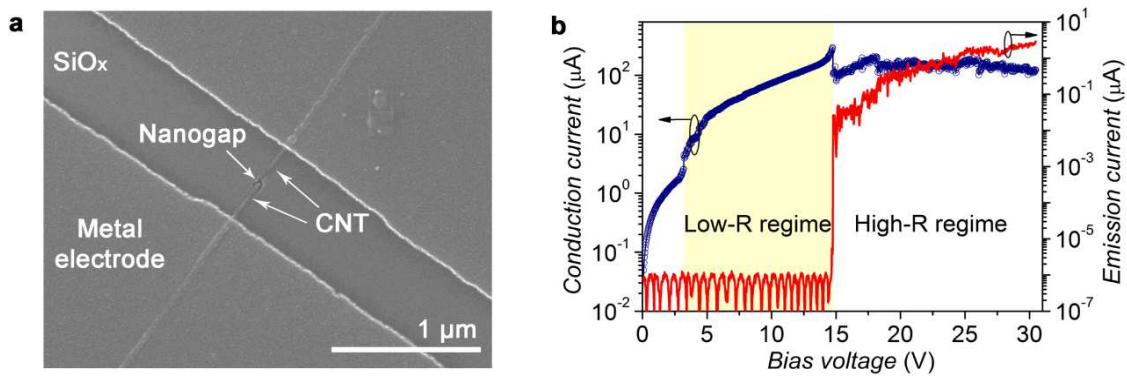

**Figure S7. EE performances of a silicon oxide EEND based on CNT electrodes.** (a) SEM image of a nanogap spaced by 18-nm-wide CNT electrodes on a silicon oxide substrate. (b) Simultaneously measured transport characteristics (blue curve) and emission characteristics (red curve) of the EEND based on CNT electrodes, which exhibited similar RS and EE performances to those of EENDs based on graphene electrodes in Figure 1 and a maximum emission current of 2.8 µA. Compared to EENDs in Figure 1, electron emission emerged at larger bias voltage of 14.7 V, which is attributed to poorer electrical contacts between CNTs and metal electrodes than those between graphene and metal electrodes.

## 8. Characterization of Graphene

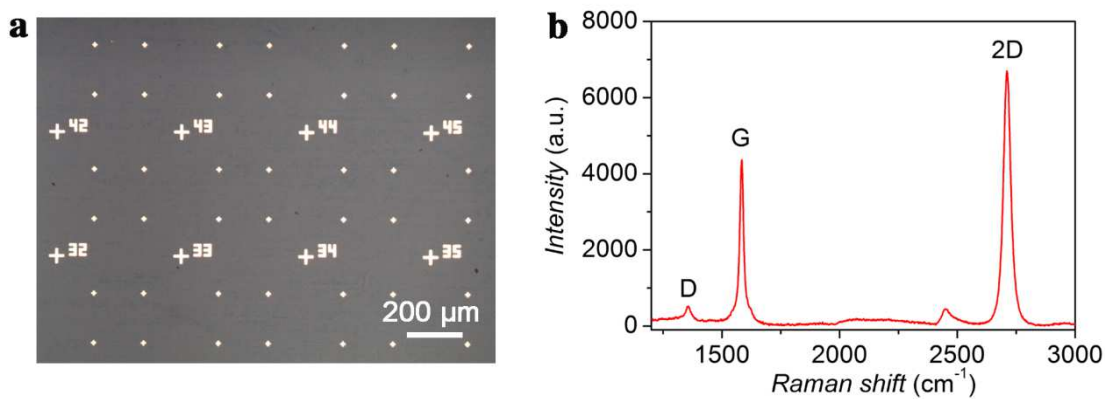

**Figure S8. Characterization of graphene.** (a) Optical microscope image of graphene after being transferred to a silicon oxide substrate with metal marks. (b) Raman spectrum of graphene in (a) with a laser wavelength of 488 nm. The ratio of 2D peak intensity to that of G peak ($I_{2D}/I_G$) lager than 1 indicates that graphene samples we used had 1 or 2 layers [7].

## 9. Electron emission measurement

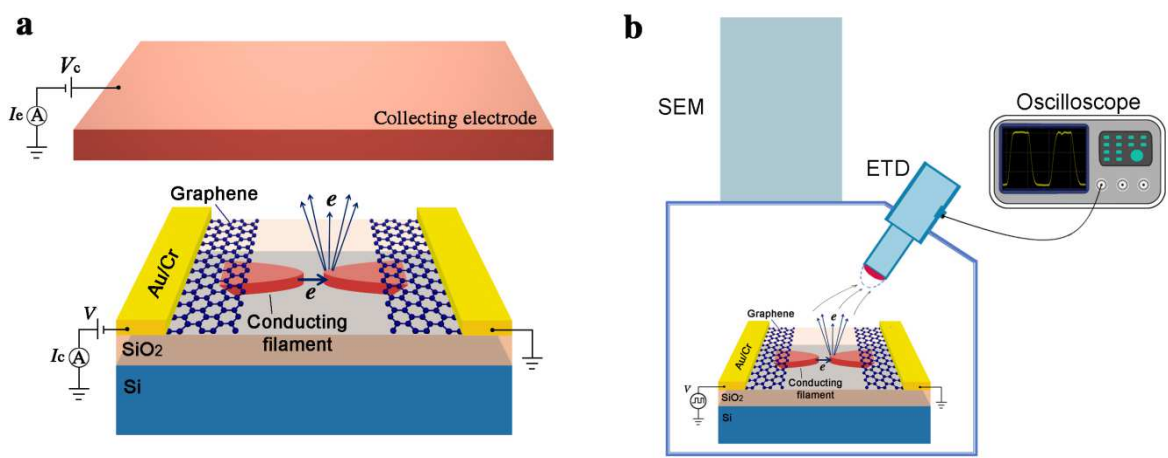

**Figure S9. Schematic setups for electron emission measurements.** (a) Schematic diagram showing the measurement of emission current from EENDs in dc mode. (b) Schematic diagram showing the measurement of the temporal response of an EEND using the Everhart-Thornley detector (ETD) in the chamber of a SEM. Note: $V$, bias voltage; $I_c$, conduction current; $V_c$, collecting voltage; $I_e$, emission current.